\documentclass[12pt,letterpaper,draftclsnofoot,onecolumn]{IEEEtran}
\usepackage{setspace}
\usepackage[margin=1in]{geometry}
\usepackage{cite}
\usepackage{amsmath,amssymb,amsfonts}
\usepackage{amsthm}
\usepackage{algorithm}
\usepackage{algpseudocode}

\usepackage{graphicx}
\usepackage{textcomp}
\usepackage{xcolor}
\usepackage{subcaption}
\usepackage{multirow}
\usepackage{booktabs}
\usepackage{ragged2e}
\usepackage{dblfloatfix}
\newcommand{\argmax}{\mathop{\mathrm{argmax}}}
\begin{document}

\title{Enhancing Exfiltration Path Analysis Using Reinforcement Learning}

\author{
        Riddam Rishu$^{a}$,
        Akshay Kakkar$^{a}$,
        Cheng Wang$^{a}$,
        Abdul Rahman$^{a}$, \\
        Christopher Redino$^{a}$,
        Dhruv Nandakumar$^{a}$,
        Tyler Cody$^{b}$,
        Ryan Clark$^{a}$, \\
        Daniel Radke$^{a}$,
        %Lanxiao Huang$^{c}$,
        %Peter Beling$^{c}$,
        Edward Bowen$^{a}$ \\
        %\small $^{b}$National Security Institute, Virginia Polytechnic University \\
         %\small $^{a}$Deloitte \& Touche Assurance \& Enterprise Risk Services India Private Limited \\
        \small $^{a}$Deloitte \& Touche LLP \\
         \small $^{b}$National Security Institute, Virginia Tech \\
        %\small $^{*}$Corresponding author: rrishu@deloitte.com \\
}
\maketitle

\begin{abstract}
Building on previous work using reinforcement learning (RL) focused on identification of exfiltration paths, this work expands the methodology to include protocol and payload considerations. The former approach to exfiltration path discovery, where reward and state are associated specifically with the determination of optimal paths,  are presented with these additional realistic characteristics to account for nuances in adversarial behavior. The paths generated are enhanced by including communication payload and protocol into the Markov decision process (MDP) in order to more realistically emulate attributes of network based exfiltration events. The proposed method will help emulate complex adversarial considerations such as the size of a payload being exported over time or the protocol on which it occurs, as is the case where threat actors steal data over long periods of time using system native ports or protocols to avoid detection. As such, practitioners will be able to improve identification of expected adversary behavior under various payload and protocol assumptions more comprehensively. 
%As in previous work, we utilize RL to support the development of a reward function based on the identification of those paths where adversaries desire reduced detection.
\end{abstract}


\section{Introduction}
In previous work \cite{cody2022discovering}, RL was employed to expose exfiltration\footnote{NIST 800-53r5 \cite{NIST800-53} states specifically that exfiltration lies within security control SC-07(10) for boundary protection to prevent unauthorized data movement (exfiltration).} (also called exfil) paths within networks through carefully engineering a reward system to account for the nodes with a network topology considered as cyber terrain \cite{conti_raymond_2018}.   Additionally,   \cite{cody2022discovering} and  \cite{maeda2021automating} did not consider the payload size or protocol preference during the exfiltration operation: these  proposed methodologies are only realistic for nominal payload sizes. Additional modelling restrictions, such as how much data is sent and at what rate the data is being moved, must be considered for large volume exfiltration operations \cite{ahmed2019real,nadler2019detection, zhan2022detecting}. These constraints reflect realistic cyber security paradigms that model adversarial activity where exfiltration operations often prefer a protocol (e.g. tunneling exfil traffic through domain name system, (DNS)) to deter detection while obfuscating intent \cite{zhang2019dns, zhan2022detecting, nadler2019detection, das2017detection, ahmed2019real}. 

The previous literature's drawbacks  are discussed in \cite{cody2022discovering} and the topic of using RL for conducting post-exploitation activities such as exfiltration is still under-studied. Previous work  \cite{maeda2021automating}, for example, employs ontological models of the agent with actions defined using common software modules. While this may be useful in some capacity, it suffers from aligning to network structure, path structure, and cyber terrain, thereby limiting its ability to anchor agents to the \emph{real} computer network. In addition, the output is not operationally interpretable where security operations center (SOC) and cyber analysts can action results \cite{nadler2019detection, das2017detection, zhan2022detecting,ahmed2019real}.

This paper presents a framework for using RL methods for discovering exfiltration paths in network models while accounting for attacker preferences in payload and protocol preferences. The key contributions of this paper are twofold:
\begin{enumerate}
    \item An approach for modeling data exfiltration on networks that accounts for choices in protocol with varying size of payload.
    \item The implementation of RL-based algorithms for discovering exfiltration paths in network models.
\end{enumerate}
The presented methodology is aligned with a focus on network structure and configuration, path analysis, and cyber terrain. Its outcomes can be directly understood as paths through networks, as is highlighted in a detailed discussion of the results. To support reproducibility, the RL solution methods, experimental design, and network model are specified in great detail. 

The remainder of this work begins with a background on the use of RL for penetration testing followed by an exploration of the methods for modeling defensive terrain and discovering exfiltration paths. Then, the experimental design is described for evaluating the proposed approach, followed by an analysis of the experimental results, and a discussion of the findings. Lastly, this paper concludes with remarks on modeling decisions, a summary of the work, and possible avenues of future research.

\section{RL and Penetration Testing}

\subsection{Reinforcement Learning}

RL is a framework where an agent learns to optimize its behaviour by interacting with its environment \cite{sutton2018reinforcement}.
A Markov decision process (MDP): $(\mathcal{S}, \mathcal{A}, P, r, \gamma)$ is often used to model the environment, where $S$ is the state space, $\mathcal{A}$ is the action space, $P: \mathcal{S} \times \mathcal{A} \rightarrow \mathcal{S}$ is the transition function, $r: \mathcal{S} \times \mathcal{A} \times \mathcal{S} \rightarrow \mathbb{R}$ is the reward function and $\gamma \in (0,1]$ is the discount factor, which determines the present value of future rewards.
The agent's behavior is characterized by its policy $\pi$, which is a probabilistic distribution over actions given a state. For deterministic policies, the action taken in state $s$ can be denoted as $\pi(s)$.
Corresponding to each time step, the agent observes a state $s_t$, on which it takes an action $a_t$ according to $\pi(a|s_t)$, and transitions to a new state $s_{t+1}$ and receives a reward $r_t=r(s_t, a_t, s_{t+1})$. The cumulative discounted reward is called the \emph{return} and is defined as $G_t=\sum_{k=0}^\infty \gamma^k r_{t+k}$.
The RL agent aims to learn an optimal policy $\pi^*$, which maximizes the expected return from each state. RL algorithms can be categorized into three groups: value function-based (or critic-only) methods, policy gradient (or actor-only) methods, and actor-critic methods.

Value function-based methods such as $Q$-learning \cite{watkins1989learning} or deep Q-networkx (DQN) \cite{mnih2015human} learn optimal policies by first estimating the optimal action-value function $Q^*(s,a)$:
\begin{align}
    Q^*(s,a)& \equiv \max_\pi Q^\pi(s,a)  \nonumber \\
    &\equiv \max_\pi\mathbb{E}_\pi\big[G_t|s_t=s, a_t=a\big],
\end{align}
which can be obtained by solving the Bellman equation:
\begin{align}
    Q^* (s,a) = \mathbb{E}_{s'} \big[ r + \gamma \max_{a'} Q^* (s',a') | s, a\big].
\end{align}
Then, an optimal policy $\pi^*$ is derived by selecting the action that yields the largest $Q$-value:
\begin{align}
    \pi^*(s) = \argmax_{a}Q^*(s,a).
\end{align}

On the other hand, policy gradient approaches focus on directly parameterizing the policy $\pi(a|s;\theta)$ and optimizing a performance measure $J(\theta)$ such as the expected return $\mathbb{E}[G_t]$ via gradient ascent.
Such methods often suffer from high variance and may result in slow learning. Thus, to reduce the variance, actor-critic methods use an estimate of the value function $V_\pi(s)\equiv \mathbb{E}_\pi[G_t|s_t=s]$ as a baseline when estimating the policy gradient $\nabla J(\theta)$ \cite{nguyen2019deep}.
The critic is responsible for learning the value function while the actor updates policy parameters by using the estimated policy gradient. In particular, the policy gradient can be estimated as 
\begin{align}
    \nabla J(\theta)  \approx \mathbb{E}\big[\nabla_{\theta}\log \pi(a_t|s_t; \theta)A_t\big],
\end{align} where $A_t=Q(s_t, a_t) - V(s_t)$ represents the \emph{advantage} of taking action $a_t$ at state $s_t$.

Policy gradient methods are prone to performance collapse as a result of large policy updates, which can be challenging to recover from because the agent will have been trained on the experience produced by bad policies. To improve training stability, Proximal Policy Optimization (PPO) \cite{schulman2017proximal} uses a clipped surrogate objective function:
\begin{align}
    \mathcal{L}(\theta) = \mathbb{E} \Big[ \min\big(\rho_t(\theta) A_t, \mathrm{clip}\big( \rho_t(\theta), 1-\epsilon, 1+ \epsilon \big) A_t\big)\Big], \label{eq:ppo_obj}
\end{align}
where $\rho_t(\theta) = \pi_\theta(a_t|s_t) / \pi_{\theta_{\mathrm{old}}} (a_t|s_t)$ is the probability ratio of the new policy over the old policy. The advantage function $A_t$  is often estimated using the generalized advantage estimation \cite{schulman2015high}, truncated after $T$ steps:
\begin{align}
    & \hat{A}_t = \delta_t + (\gamma\lambda) \delta_{t+1}+\cdots + (\gamma\lambda)^{T-t+1}\delta_{T-1},\\
    & \mathrm{where\;} \delta_t = r_t + \gamma V(s_{t+1}) - V(s_t).
\end{align}
To support exploration, an entropy bonus $\beta H(\theta)$ is often added to the objective function \eqref{eq:ppo_obj}, where $\beta$ is a coefficient.

\subsection{RL applications in penetration testing}
Deep RL has been applied to cybersecurity broadly \cite{nguyen2019deep}, but only recently it has been employed as a tool for penetration testing \cite{ghanem2018reinforcement, schwartz2019autonomous, ghanem2020reinforcement, chaudhary2020automated, yousefi2018reinforcement, chowdhary2020autonomous, hu2020automated, gangupantulu2021using}. There are a number of different approaches, but most only consider privilege escalation on a target host as the learning task. Gangupantulu et. al proposed to use concepts of cyber terrain to help enrich task design and reward shaping \cite{gangupantulu2021using}. This concept spurred the development of several task-specific uses of RL for penetration testing, including crown jewel analysis \cite{gangupantulu2021crown}, discovering exfiltration paths \cite{cody2022discovering}, and exposing surveillance detection routes \cite{huang2022exposing}.

As with Gangupantulu et al.\cite{gangupantulu2021crown}, the presented RL approach here solves a more complex task and acts as a focused tool for cyber operators to increase the effectiveness of operator workflow in penetration testing.
RL for penetration testing has made frequent use of DQN \cite{schwartz2019autonomous, chowdhary2020autonomous, hu2020automated, gangupantulu2021using, gangupantulu2021crown}. As an alternative, Nguyen et al. proposed an RL-based approach that makes use of two agents: one for iteratively scanning the network to build a structural model and another for exploiting the constructed model \cite{nguyen2021proposal}. 
In this study, Nguyen et al.'s double agent architecture is combined with the PPO algorithm to train the RL agents.

\section{Methods}
In this section, we present the details of the exfiltration model,  the protocol-based path selection criteria, and the complete RL formulation. 
%It is worth noting that while the development of our model necessitates certain assumptions that may not precisely mirror reality, it is underpinned by a data-driven approach. The inputs to the model, derived from scan data (such as Nmap or Nessus), offer flexibility for future refinements to enhance its alignment with real-world conditions.
While the model incorporates several assumptions, it's fundamentally based on a data-driven approach using scan data. This reliance on scan data not only ensures empirical robustness but also permits iterative refinements, as newer or more comprehensive data become available, to progressively approach a more accurate representation of reality.

\subsection{Exfiltration Simulation Overview}

The approach proposed here expands on previous  models for data exfiltration in that it can model paths for different payload sizes. The exfiltration campaign is modeled based on three tasks consisting of (i) Connection, (ii) Path Selection and (iii) Exfiltration. The attacker initially attempts to gain control of some of the known target hosts which are externally connected via an internet connection to serve as the point of exfiltration. Once control of the target host is gained, an exfiltration path is selected based on the preferences for an exfiltration protocol. The attacker then tries to exfiltrate data packets from the compromised host. The three tasks are designed to function so that if an attacker discovers a new exfiltration host, the path selection module determines whether a better path exists and adjusts the exfiltration path accordingly.

The agent explores the network and gathers information on neighboring hosts by taking the subnet scan action. In order for the scan to be successful, the agent must first gain access to the underlying host, which can be achieved by executing an exploit action. Multiple exploits may exist for a given machine, with each targeting a specific Common Vulnerabilities and Exposures (CVE) vulnerability. 

Once a foothold is gained on a new host, the agent updates candidate exfiltration paths that consist of each of the compromised hosts. 
It will then decide which path is preferred to carry out the exfiltration based on the predetermined Exfiltration protocol strategy. If another target is captured later, the path selection task evaluates the new paths available and, if a new preferred path is discovered, the path is updated and the payload is reset to its original value.

After identifying the preferred exfiltration path, the agent can start sending parts of the payload to the exit node. The task is completed if the entire payload is uploaded from the initial node. In order to evade firewall detection, the agent should avoid frequent and large uploads. To hide its activity, the agent may take a sleep action that simply does nothing for a period of time. 

\subsection{Network Firewalls}
As in \cite{wang2023discovering}, any exfiltration traffic will be monitored by network firewalls, which are placed between each of the subnets and the public Internet. Upon detection of unusual traffic patterns, the administrator will be alerted and an emergency firewall update will be conducted. Examples of suspicious activities include the following:
\begin{itemize}
    \item the total egress volume exceeds \texttt{max\_upload\_volume};
    \item the total active time surpasses \texttt{max\_upload\_time}.
\end{itemize}
Table \ref{tab:firewall} lists the values of firewall-related parameters used in the experiments.
 \begin{table}[h]
     \centering
     \begin{tabular}{|c|r|} \hline
      \textbf{Firewall Parameter} &  \textbf{Value}\\ \hline
         max\_upload\_volume (MB) & 5000 \\
         max\_upload\_time (minutes) & 4 \\
         update\_frequency (hours) & 24\\ \hline
     \end{tabular}
     \caption{Firewall Parameters.}
     \label{tab:firewall}
 \end{table}

Firewalls are also updated periodically. In particular, a wall-clock is introduced to simulate the real time of an attack campaign. Different actions will increase the clock time by different amounts depending on the their complexity.
Both the regular update and the emergency update will patch the vulnerabilities and block the outbound traffic from the compromised hosts.

\subsection{Protocol-Based Path Selection}

Exfiltration activities within attacker campaigns are typically carried out by exploiting a common protocol as these are deemed generally safer and less likely to be detected by security monitoring. Standard protocols, such as 
Hypertext Transfer Protocol Secure (HTTPS), are often used to carry out data exfiltration. By using common protocols used by enterprise applications, it's more likely these protocols are available. It's also more likely these protocols are not monitored as closely by security detection methods. As an example, by using the same protocol used by databases to backup their data to cloud services, attackers emulate the database backup expected by security rules and do not raise alerts in monitoring systems.

Path selection is determined by maximizing the utilization of this protocol across as many hosts in an exfiltration network path as possible. This is not always the shortest path. A path maximizing the use of the chosen protocol is often more advantageous, even when this path touches more nodes in the victims network. The path selection algorithm accounts for contingencies when end-to-end use of the designated exfiltration protocol is unavailable. The algorithm prioritizes finding a complete path using the given protocol over a shortest path possible. The next level of criterion considered are the length of the path and rewards accumulated. If multiple paths are identified with the same exfiltration protocol coverage, the shortest path will be prioritized. When no complete path can be created using the exfiltration protocol, the algorithm searches for the shortest path exposed to the maximum use of the protocol. The reward function calculates the highest rewarded path using existing reward mechanisms, shortest number of hosts, and maximum use of the exfiltration protocol.

\subsection{Reinforcement Learning Formulation}
\subsubsection{State Space}
The state has the following features for every host:
\begin{itemize}
    \item Address, 
    \item Operating system,
    \item Services and processes,
    \item Discovery value and status, 
    \item Infection value and status,
    \item Access level information.
\end{itemize}
Host's address is denoted by its subnet ID and local ID. The operating system, service and process features  have a value of one of if they are present at the host and zero otherwise. Similarly, the discovery and infection status are one if the host is discovered or compromised and zero otherwise. The discovery and infection values represent the reward for successfully discovering and compromising a host, respectively. Additional features are defined for target hosts:
\begin{itemize}
    \item Connection status,
    \item Time since infection,
    \item Remaining payload size,
\end{itemize}
The connection status can be connected, not connected, or isolated (i.e., blocked by firewalls). The time since infection is measured by the wall-clock rather than time steps. Finally, the remaining payload size indicates how much left to upload. The exfiltration task is complete when the remaining payload size becomes zero.

\subsubsection{Action Space}
There are four types of actions for the RL agent: \emph{subnet scan}, \emph{exploit}, \emph{upload}, and \emph{sleep}.
Each action requires specification of a target host, except for the sleep action, which simply does nothing for a given period of time. Multiple exploits targeting at different vulnerabilities may be available for a given host. Two uploading actions with different speed are available at each target - one with a rate of 100MB/s and another with rate 1MB/s.

Clock-time increases differently based on the action's result and complexity. Table \ref{tab:actions} lists the assigned clock time for each action. For not applicable actions, such as performing as subnet scan without access to the underlying host, the clock time will only move forward by one second.

\begin{table}[h]
    \centering
    \caption{List of actions.}
    \begin{tabular}{|c|r|} \hline
       \textbf{Action Type}  & \textbf{Time} \\ \hline
       Subnet Scan & 30\\
       Exploit  & 10 \\
       Upload  & 10\\
       Sleep  & 60\\ \hline
    \end{tabular}
    \label{tab:actions}
\end{table}

\subsubsection{Reward Function}

The reward function consists of a positive value for achieving sub-goals such as discovering or exploiting a host and a negative value that accounts for the action's cost. An action with higher cost is more likely to trigger the defense terrain. Specifically, we follow the approach in \cite{cody2022discovering} and assign action' cost based on the services running on the target system. The idea is that even though the adversaries may not know the exact defense mechanism or strength, they can still infer the presence of defense based on the host's service information. In particular, we categorized the services into three groups, high-risk, medium-risk, and low-risk. The actual cost of an action then depends on its type (scan, exploit, or upload) and the target's service profile.

\begin{table}[h]
    \centering
    \caption{List of rewards.}
    \begin{tabular}{|c|r|} \hline
        \textbf{Reward Type} & \textbf{Value} \\ \hline
        Discovery &  1000\\ 
        Exploit & 1000\\ 
        Exfiltration Protocol Path & 1000\\ 
        Upload (per unit) & 0.1 \\ 
        Upload (bonus) & 10000 \\ \hline
    \end{tabular}
    \label{tab:rewards}
\end{table}

Rewards are given based on how much of the exfiltration path chosen is covered by the exfiltration protocol, For example if out of the 6 hosts in the exfiltration path 3 hosts have exfiltration protocol running then 50 percent of the reward configured will be given to the agent. The agent receives positive reward on uploading a partial payload from the infected host, upon finishing sending the entire payload, the agent is given a large bonus reward.  However, if exfiltration is detected by network firewalls, then the agent will receive a penalty equal to the total accumulated rewards gained on the originating host and the host will be isolated. That is, the agent will lose all rewards from discovery, infection and partial uploads.
Table \ref{tab:rewards} lists rewards used in this study.

\section{Experiments}
In this section we present the experiment details and the results, and discuss key characteristics of the attack paths learned by the RL agent.

\subsection{Network Description}

We have designed two experiment networks. 
The first experiment network has 10 subnets and a total of 56 hosts. Each subnet contains between 3 and 12 hosts. The attacker agent is assumed to have gained an initial foothold on host (8, 2) in subnet 8, which is not directly connected to the Internet. One particular machine (2, 0) from subnet 2 is designated as the exfiltration host. Subnet 2 is directly accessible from the internet whereas, other subnets are private and are not directly accessible from the Internet. The exfiltration host has Dynamic Host Configuration Protocol Server (DHCPS) running as a service, which is chosen as the Exfiltration Protocol. 
The second experiment network has 101 subnets and a total of 1444 hosts. This network is remarkably bigger than the one used previously. Each subnet contains between 3 and 50 hosts. The attacker agent is assumed to have gained an initial foothold on host (44, 5) in subnet 44, which is not directly connected to the Internet. A host connected to the internet (5, 10) from subnet 5 is designated as exfiltration host. The exfiltration host has running HTTPS service, which is chosen as the Exfiltration Protocol.

\subsection{Training Details}
\begin{table}[h]
    \centering
    \caption{List of hyperparameters.}
    \begin{tabular}{|l|c|} \hline
        \textbf{Hyperparameter} & \textbf{Value} \\ \hline
        Critic learning rate ($\alpha_w$) & 0.0003\\ 
        Actor earning rate ($\alpha_\theta$) & 0.0003\\ 
        Discount factor ($\gamma$) & 0.99\\
        Horizon (T) & 2048\\
        Minibatch size & 32\\ 
        Epochs & 5\\ 
        GAE parameter ($\lambda$) & 0.95 \\ 
        Clipping parameter ($\epsilon$) & 0.2 \\
        Entropy coefficient ($\beta$) & 0.02 \\ \hline
    \end{tabular}
    \label{tab:hyperparameters}
\end{table}
% #means the value hasn't been changed.

The RL agent is trained in an episodic fashion for both the networks using the well-known PPO algorithm. An episode ends when the initial host either completes sending payload to the exfiltration host or is isolated by firewalls. The target payload is set to be 10,000MB. Both the actor and the critic are approximated by a two-layer feed-forward neural network, where the first layer has 64 neurons, and the second layer has 32 neurons. Other key hyperparameters are listed in Table \ref{tab:hyperparameters}. 
For the first network the RL agent is trained for 800 episodes and for the second network RL agent is trained for 1000 episodes.

\section{Results}

\begin{figure}[h]
\begin{subfigure}{.5\textwidth}
  \centering
  % include first image
  \includegraphics[width=.8\linewidth]{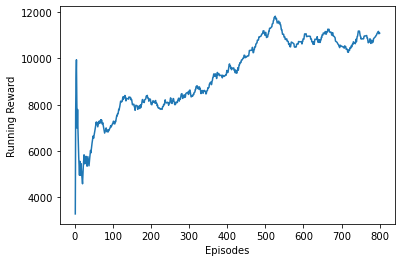}  
  \caption{}
  \label{fig:ip50_reward}
\end{subfigure}
\begin{subfigure}{.5\textwidth}
  \centering
  % include second image
  \includegraphics[width=.8\linewidth]{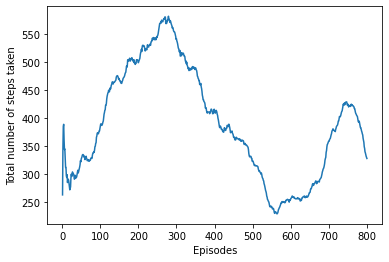} 
  \caption{}
  \label{fig:ip50_steps}
\end{subfigure}
\caption{Average episode reward (left panel) and length (right panel) from the first network.}
\label{fig:plots_Network1}
\end{figure}

\begin{figure}[h]
\begin{subfigure}{.5\textwidth}
  \centering
  % include second image
  \includegraphics[width=.8\linewidth]{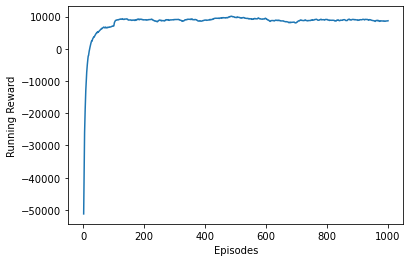}  
  \caption{}
  \label{fig:ip1500_reward}
\end{subfigure}
\begin{subfigure}{.5\textwidth}
  \centering
  % include second image
  \includegraphics[width=.8\linewidth]{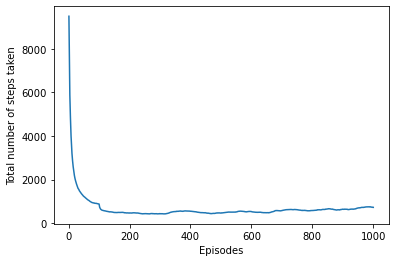}  
  \caption{}
  \label{fig:ip1500_steps}
\end{subfigure}
\caption{Average episode reward (left panel) and length (right panel) from the second network.}
\label{fig:plots_Network2}
\end{figure}

For the first network, episode rewards over training runs are presented in Fig. \ref{fig:ip50_reward} and episode length in Fig. \ref{fig:ip50_steps}, and for the second network, episode rewards over training runs are presented in Fig. \ref{fig:ip1500_reward} and episode length in Fig. \ref{fig:ip1500_steps}. 
Training is observed to be stable for both networks, and the RL policy converges in 800 episodes for the first network and in 1000 episodes for the second network. 
Fig. \ref{fig:ip50_reward} shows that the sum of rewards in an episode for the first network steadily increases to almost 12,000, and Fig. \ref{fig:ip1500_reward} shows that the sum of rewards in an episode for second network steadily increases to a little more than 10,000. During the same intervals, the episode length gradually decreases for both simulations. This suggests that as training goes on, the RL agent completes the attack task more efficiently and takes fewer random actions.

% subfigure for RL plots

Table \ref{tab:statsnet1} reports statistics on the length and rewards from the generated attack paths for the first network. On average, the RL agent finishes the task in 389 steps and receives a total reward of 9298. Table \ref{tab:statsnet2} reports statistics on the length and rewards from the generated attack paths for the second network. On average, the RL agent finishes the task in 670 steps and receives a total reward of 8252.

\begin{table}[h]
    \centering
    \caption{Summary statistics of the generated attack paths for First Network}
    \begin{tabular}{|l|c|c|} \hline
        \textbf{} & \textbf{Steps} & \textbf{Rewards}\\ \hline
        Mean & 389 & 9298\\ 
        Std & 98 & 1618\\ 
        Min & 229 & 3292\\
        Max & 582 & 11814\\ \hline
    \end{tabular}
    \label{tab:statsnet1}
\end{table}

\begin{table}[h]
    \centering
    \caption{Summary statistics of the generated attack paths for Second Network}
    \begin{tabular}{|l|c|c|} \hline
        \textbf{} & \textbf{Steps} & \textbf{Rewards}\\ \hline
        Mean & 670 & 8252\\ 
        Std & 608 & 3898\\ 
        Min & 427 & -51249\\
        Max & 9493 & 10103\\ \hline
    \end{tabular}
    \label{tab:statsnet2}
\end{table}

Due to the stochastic nature of the learned policy, the RL agent may take some unnecessary or redundant actions such as exploiting unimportant hosts or subnet scans. After pruning the output trajectory,  key steps in the attack for the simulation of the first network can be identified as shown in Table \ref{tab:stepsfornet1}.
\begin{table}[h]
    \centering
    \caption{List of main steps taken by the RL agent for First Network}
    \begin{tabular}{|l|c|} \hline
        \textbf{Action} & \textbf{Target} \\ \hline
        Subnet Scan & (8, 2)\\ 
        Exploit & (4, 2)\\ 
        Subnet Scan & (4, 2)\\
        Exploit & (2, 0)\\
        Exploit & (6, 0)\\ 
        Subnet Scan & (6, 0)\\ 
        Exploit & (5, 1) \\ 
        Upload(10 MB) & (8, 2) \\
        Upload(1000 MB) & (8, 2) \\
        Sleep(NoOp) & - \\ \hline
    \end{tabular}
    \label{tab:stepsfornet1}
\end{table}

For the first network, the agent gains a foothold on host (8, 2) in subnet 8, from which it triggers a subnet scan which leads to the discovery of other hosts in the same subnet and in the connected subnets, subnet 4 and subnet 6. The agent then exploits the host (4, 2) in subnet 4 and it is chosen as a host for further exploitation to make an exfiltration path. A subnet scan is triggered from the host (4, 2), which discovers the hosts present in connected subnets i.e., subnet 2 and ultimately discovers the target or exfiltration host (2, 0), which is then compromised to forge an exfiltration path i.e., $(8,2) \to (4,2) \to (2,0)$. In search of availability of better paths, agent exploits the host (6, 0) in subnet 6, and triggers a subnet scan from that host, discovering hosts on connected subnet i.e., subnet 5. This scan discovers host (5, 1) in subnet 5 and is later exploited to forge another exfiltration path i.e., $(8,2) \to (6,0) \to (5,1) \to (2,0)$.

\begin{figure}[h]
    \centering
    \includegraphics[width=0.95\textwidth]{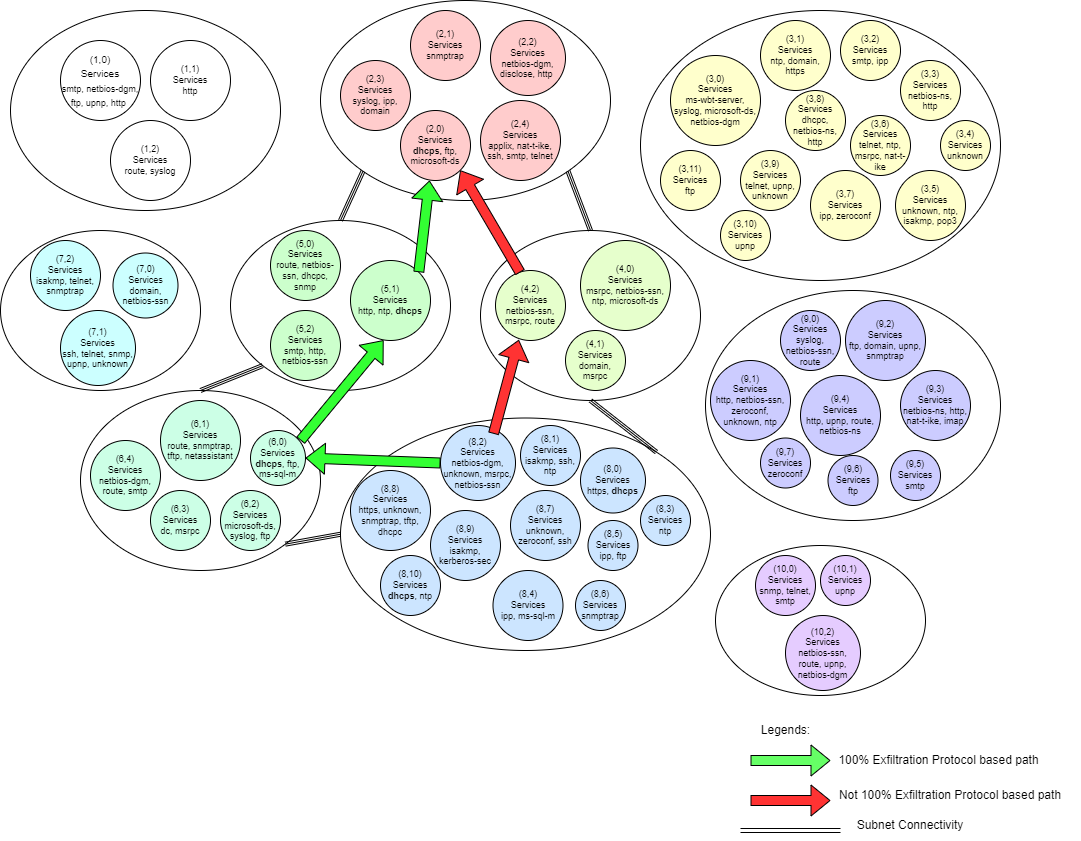}
    \caption{
    Network diagram showing the two exfiltration paths found. The green path is preferred over the read one as it utilizes the same protocol (i.e., DHCP).}
    \label{fig:pathsnet1}
\end{figure}

The path explored earlier i.e., $(8,2) \to (4,2) \to (2,0)$ is not a complete exfiltration protocol-based path, since there is no DHCP (exfiltration protocol) service running on host (4, 2) as shown in Fig. 
\ref{fig:pathsnet1}. However, the second path discovered i.e., $(8,2) \to (6,0) \to (5,1) \to (2,0)$ is a complete exfiltration protocol-based path since the same service (i.e., DHCP) is running on both hosts (6, 0) and (5, 1) as shown in Fig. 
\ref{fig:pathsnet1}. Noticeably, the agent chooses the second path over first path to upload payload as it is 100 percent protocol-based path and is the optimal path, even though the first path discovered is shorter in length.

For the second network, the agent has a foothold over the host (44, 5) in subnet 44. Upon performing various subnet scans and exploits, the agent gets a hold over the host (24, 18), and ultimately discovers and exploits the target or exfiltration host (5, 10). This led to development of exfiltration path i.e., $(44,5) \to (24,18) \to (5,10)$. 
The host (24, 18) has HTTPS service running on it, hence the path forged is a complete protocol-based path. The capability of the agent to forge a 100 percent protocol-based path over such a big network indicates that the model is scalable as well.

For both networks the agent takes appropriate sleep actions in between the upload actions so that there is no unusual traffic pattern and cyber defenses are not triggered.

The agent found paths in both networks that utilize a single network protocol. In real-life scenarios, attackers try to use a single protocol to avoid increasing attack complexity and reduce the risks of inconsistencies or errors, which can lead to a greater possibility of detection. Choosing to exfil data using existing network protocols that the network defenses (firewalls, IDS) know about also reduces the risk of discovery by traffic anomaly detection algorithms. Using standard protocols for exfiltration while considering traffic timing and volume replicates previously documented Tactics, Techniques, and Procedures (TTP)s\cite{mitre-attack}.

While novel exfiltration methods that use non-standard protocols exist,Domain Name Service (DNS), Network Time Protocol (NTP), or Internet Control Message Protocol (ICMP), they typically require complex setup for execution \cite{zhang2019dns}. They also are usually more closely monitored by defensive measures for volume and anomalous behaviors than standard protocols due to their usage in previous exfiltration operations \cite{zhang2019dns}. Data exfiltration requires more network volume and can be more stealthily sent over less strictly monitored or eccentric channels \cite{ml-detect-data-exfil}. 

\section{Conclusion}
The current gap within the cybersecurity industry involves contextualizing and quantitatively prioritizing the efficacy of deployed security controls to enable sense-making for security practitioners and network defenders. In this paper, we address this gap through applying RL for exfiltration path analysis enhanced by integrating protocol and payload considerations. Our work demonstrates that an RL agent can effectively find an exfiltration path with maximum exfiltration protocol coverage and can perform exfiltration using this preferred path without being detected by security infrastructure (i.e., firewalls). Our results identify optimal paths that provide insights for operators, analysts, and defenders to evaluate the value of currently deployed security controls which influence (i.e., isolate or eliminate) the connections within the path. As a result, the operations community can utilize this data to formulate task lists for securing enterprise networks.  

This RL approach identified the most likely hosts and services used when exfiltrating data while capturing variable metrics used in network risk assessments. The strength of this approach was validated through identification of intentional network misconfigurations that mimic real-world vulnerabilities.  In future work we consider expanding the risk formalism to increase its sophistication and maturity, which will drive increased applicability and relevance.

% \section{Acknowledgements}

% This work was funded by Deloitte's Artificial Intelligence Center of Excellence (AICoE) within Deloitte Advisory's AI Strategic Growth Offering (SGO) in collaboration with Deloitte Advisory's Cyber SGO and the Hume Center for National Security and Technology's Intelligent Systems Lab at the Virginia Polytechnic Institute and State University.

\bibliographystyle{IEEEtran}
\bibliography{ref}

\end{document}